\documentclass[aps,preprintnumbers,amsmath,amssymb,twocolumn,showpacs]{revtex4}

\usepackage{amsmath,amssymb,mathrsfs}
\usepackage{graphicx}

\def\vx{{\mathbf x}}

\newcommand{\pri}{\!\,^{'}}

\begin{document}

\title{Attractive Casimir Forces in a Closed Geometry}

\author{M.~P.~Hertzberg,$^{1,2}$ R.~L.~Jaffe,$^{1,2}$ M.~Kardar,$^2$ and A.~Scardicchio$^{1,2}$}
\affiliation{$^1$Center for Theoretical Physics, Laboratory for Nuclear Science, and \\ $^2$Department of Physics\\
Massachusetts Institute of Technology, Cambridge, MA 02139, USA}


\begin{abstract}
We study the Casimir force acting on a  conducting piston with arbitrary cross section.  
We  find the exact solution for a rectangular cross section and the first three terms in 
the asymptotic expansion for small height to width ratio when the cross section is arbitrary.  
Though weakened by the presence of the walls, the Casimir force turns out to be always attractive. 
Claims of repulsive Casimir forces for related configurations, like the cube, are invalidated by cutoff dependence.
\end{abstract}
\vspace*{-\bigskipamount} \preprint{MIT-CTP-3677} 

\maketitle
 
In 1948 Casimir predicted a force between conducting plates due to
quantum electromagnetic fluctuations\cite{Casimir}.  Casimir
forces have now been measured precisely\cite{exp},
and agree with theory to a few percent accuracy.  As
miniaturization continues, static friction due to the Casimir
attraction between components in microelectromechanical systems
(MEMS) has become a problem of increasing concern\cite{ChanandSerry}.  
Since the Casimir force is strongly geometry dependent\cite{GolesandEmig}, 
the question arises whether it can be made repulsive by arranging conductors appropriately.   
It has been claimed that the parallelepiped of sides $a,b,b$ has positive Casimir energy for 
aspect ratio in the range $0.408<a/b<3.48$ \cite{group,Wolfram}.  
If so, a repulsive force would occur for $a/b>0.785$, as $a$ is varied at fixed $b$. 

However this example, and indeed all claims of repulsive Casimir forces, require 
elastic deformations of single bodies (a rectangle or a circle in 2-dimensions, a parallelepiped, 
a cylinder or a sphere in 3-dimensions, typically) 
treated as perfect, often infinitely thin conductors.  It has been shown that, when the 
conductors are modeled more realistically, the Casimir energies  
associated with deformation 
actually depend strongly on material properties such as the plasma frequency and skin depth\cite{realistic}, 
and diverge in the perfect metal limit where these cutoffs are ignored. In contrast, 
the forces between rigid bodies remain finite in that limit.   Recently Cavalcanti\cite{Caval} 
introduced a modification of the rectangle --- the ``Casimir piston'' --- that is 
demonstrably free of cutoff dependence\cite{divergence}.   The 2-dimensional piston studied in Ref.~\cite{Caval} 
consists of a single rectangle divided in two by a partition  (the ``piston'').  Cavalcanti calculated the Casimir 
force on the {piston} due to fluctuations of a scalar field obeying Dirichlet boundary conditions 
on all surfaces.  He found that the force on the {piston} is always attractive, although substantially 
weakened with respect to parallel lines.

In this Letter and the coming paper\cite{following} we consider the 3-dimensional piston, for both  
scalar and electromagnetic (EM) fields.  
We keep careful track of possible cutoff dependences and show that they 
cancel for the piston configuration\cite{divergence}. 
First we give the exact result for pistons of 
rectangular cross section, expanding the result in powers of the {piston}-base 
separation, $a$, and identify the terms with specific optical paths\cite{Scard}. 
Next we consider pistons of \emph{arbitrary} cross section, where an expansion for 
small $a$ can be derived.  Finally we show that the force on the {piston}  
is always attractive for a rectangular cross section and argue that this is true for any cross section.  

\begin{figure}[h]
\begin{center}
\scalebox{0.73}{\includegraphics{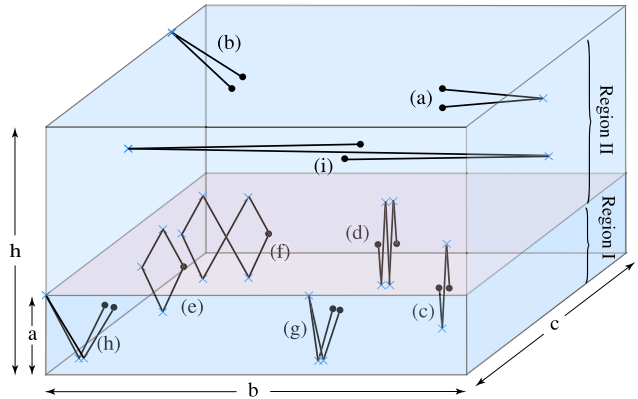}}
\caption{\label{2dnew} Optical paths in a 3-dimensional box of size
$h\times b\times c$ with a partition at height $a$.  The origin and end points (which actually coincide) of some paths are 
separated so the path can be visualized.  See the text for further description.}
\end{center}
\end{figure}

The Casimir energy for a domain ${\cal D}$ can be defined by ($\hbar=c=1$),
$E(\Lambda)=\frac{1}{2}\sum_{m}^{\Lambda}\omega_{m}({\cal D})-E_0$,
where $\omega_m({\cal D})$ are eigenfrequencies in the domain ${\cal D}$, $E_0$ is the energy of the vacuum, and
$\Lambda$ is a cutoff. The Casimir energy can be related to the Green's function, 
$G(\vx',\vx,\mathcal{E})$, (here $\mathcal{E}=\omega^{2})$ that obeys $(-\nabla'^2-\mathcal{E})G(\vx',\vx,\mathcal{E})=\delta^3(\vx'-\vx)$ and the 
appropriate boundary condition, through: $E(\Lambda)= \frac{1}{2}\int^{\Lambda^2}\!\! d\mathcal{E} \sqrt{\mathcal{E}}\rho(\mathcal{E})$ where 
$\rho(\mathcal{E})\propto \Im\int \!d\vx\,G(\vx,\vx,\mathcal{E})$.  We find it useful to expand $G(\vx',\vx,\mathcal{E})$ in a series of ``optical paths'' 
as in Ref.~\cite{Scard}.  The sum over optical paths isolates cutoff dependences, highlights the most 
important contributions, and \emph{is exact for rectilinear geometries}, provided paths touching edges and corners 
are properly included\cite{following}.  
(We explicitly checked the results for the parallelepiped by other methods.)
The Green's function is given by a {discrete} sum over straight line paths, {$p_{r}$ (some of which are shown in Fig.~\ref{2dnew}),
from $\vx$ to $\vx'$ (of length $l_{p_r}(\vx',\vx)$), undergoing specular reflection $r$ times from surfaces ({\it eg.\/} Fig.~\ref{2dnew}(a), (c), and (f)), 
edges (Fig.~\ref{2dnew}(b) and (g)), and corners (Fig.~\ref{2dnew}(h)).  Each path is weighted by a phase, $\phi(p_{r})=\eta^{n_{s}+n_{c}}$,} where $n_{s}$
and $n_{c}$ 
are the number of surface  and corner reflections on the path and $\eta=-1(+1)$ for Dirichlet (Neumann) boundary conditions (BC).  
The \emph{cutoff independent}, $a$-dependent part of the Casimir energy in a rectilinear geometry is
\begin{equation}
\widetilde{E}_{\eta} =-\frac{1}{2\pi^2}\sum_{r\ge 2}\phi(p_{r})\!\int_{\mathcal{D}}d\vx\frac{1}{\left[l_{{p}_r}({\bf
x})^{4}\right]},
\label{enfin}
\end{equation}
here $l_{p_r}({\bf x})$ is the coincidence limit (${\bf x}'\to{\bf x}$) of path lengths.
The \emph{cutoff dependent} contributions (in the sense of \cite{divergence}) are isolated in the one-reflection term, 
and has been omitted from   eq.~(\ref{enfin}).  Some one-reflection paths are shown in Fig.~\ref{2dnew}(a) and \ref{2dnew}(b).

From considering the cutoff dependent contributions of one reflection paths in Region I (see {Fig.~\ref{2dnew}}), 
we obtain\cite{following} the following Casimir energy as $\Lambda\to\infty$
\begin{equation}
E^{I}_{\eta}(\Lambda)=\frac{\eta}{8\pi}S\Lambda^3+\frac{1}{32\pi}L\Lambda^2+\widetilde{E}^I_{\eta}~,
\label{struct}
\end{equation}
where $S=2(ab+bc+ac)$ is the surface area and $L=4(a+b+c)$ is the perimeter length. 
The cutoff dependent terms agree exactly with the predictions of Balian and Bloch's theory of the density of 
states in bounded domains\cite{Balian:1970fw}.  
For the geometries considered in Refs.~\cite{group,Wolfram} these divergences are not cancelled by other 
contributions {\it eg.\/} from the exterior (discussed below for the electromagnetic case), but are instead dropped as part of a physically 
unmotivated ``renormalization'' process.   In reality, deformations of a single parallelepiped that 
change $S$ or $L$ cannot be calculated independently of the material properties of the metal that 
determine the cutoff. However, when the contribution of the upper part of the piston (Region II) is included, 
the $a$-dependence disappears from the
cutoff dependent terms, leaving a cutoff independent force in the limit $\Lambda\to\infty$.  
For a scalar field obeying Dirichlet ($\eta=-1$) or Neumann ($\eta=+1$) 
BC, and setting $b=c$ for convenience, the force is
\begin{eqnarray}
\label{parallel}
F_{\eta}&=&-\frac{\partial}{\partial
a}\left(\widetilde{E}_{\eta}(a,b,b)+\widetilde{E}_{\eta}(h-a,b,b)\right)\nonumber\\
 &=&-\frac{3\zeta(4)}{16\pi^2 a^4}A-\eta\frac{\zeta(3)}{32\pi a^3}P-\frac{\zeta(2)}{16\pi a^2}\nonumber\\
&&-\frac{J_{\eta}}{32\pi^2 A}+\mathcal{O}\!\left(\frac{b e^{-2\pi
b/a}}{a^3}\right), \,\,\mbox{as}\,\,a\to 0~,
\end{eqnarray}
where $A=b^2$ and $P=4b$ are the area and perimeter of the base, respectively.
We have expanded in $a\ll b=c$ and taken the total height of the box $h\to \infty$, the configuration most relevant to potential experiments.  
Complete results will be presented in Ref.~\cite{following}.  
Here  $J_{\eta}\equiv Z_2(1,1;4)+2\pi\eta\zeta(3)$, where 
$Z_{m}(x_{1},x_{2},\ldots,x_{m};n)$ is the Epstein Zeta function\cite{Wolfram}
and $\zeta(n)$ is the Riemann Zeta function, 
such that $J_{-1}\approx-1.5259$ and $J_{+1}\approx 13.579$.

The terms in eq.~(\ref{parallel}) can be understood through the optical paths associated with them.  
The most important contributions have the shortest path length, $l_{p_r}(\vx)$.  
The leading contribution, $\sim 1/a^{4}$, comes from paths that reflect only off both the base and the {piston}, beginning with the 2 reflection path, Fig.~\ref{2dnew}(c), and continuing through 4 (Fig.~\ref{2dnew}(d)), 6, 8, $\cdots$ reflections.  They sum up to the well-known parallel plate result, the first term in eq.~(\ref{parallel}).  The first correction, $\sim 1/a^{3}$, comes from closed but not periodic paths that additionally reflect once from a wall of the piston, beginning with the three reflection path,  Fig.~\ref{2dnew}(e), and continuing through 5 (Fig.~\ref{2dnew}(f)), 7, {\it etc.\/} Their sum is related to the Casimir force between lines in two dimensions\cite{following}.  The next correction, $\sim 1/a^{2}$, comes from paths that reflect off edges, beginning with the three reflection path, Fig.~\ref{2dnew}(g).  It is related to the Casimir force between points on a line.  The $a$-independent fourth term in eq.~(\ref{parallel}) arises from paths in Region II that
do not reflect from the {piston} (see Fig.~\ref{2dnew}(i)). 
  Note that the constant term gives a repulsive (attractive)
contribution for Dirichlet (Neumann) BC.
The remaining terms are exponentially small as $a\to 0$ due to an effective mass for the corresponding paths,
 which we explain in detail later.

We now turn to the case of electromagnetism, which is of most physical interest.  To construct the Casimir energy, we must determine the frequency spectrum. Inside the cavity ${\bf E}$ and ${\bf B}$
obey the wave equation,  $\nabla^2({\bf E,B})-\partial_{t}^{2} ({\bf E},{\bf B})={\bf 0}$, supplemented by conducting boundary conditions at the walls, ${\bf n\times E} ={\bf 0}$ and ${\bf n\cdot B} =0$. In Region I, the solution  for ${\bf E}$ is,
\begin{eqnarray}
{\bf E}_{x} =  A_{x}\cos(n\pi x/a)\sin(m\pi y/b)\sin(l\pi
z/c)e^{-i\omega t}~,
\end{eqnarray}
and similarly for ${\bf E}_y,{\bf E}_z$.  Here $A_{x,y,z}$ are constants and $n,m,l$ are integers. 
The magnetic field is obtained from ${\bf B}=-i\nabla\times {\bf E}/\omega$.
The transversality of ${\bf E}$, $\nabla\cdot{\bf E}=0$, restricts the degeneracy of the modes:  If all integers are non-zero,
there is a twofold degeneracy. If one of the integers is zero, there is a single solution. If two or more of the integers are zero, the mode is forbidden.
Taking note of this, we may express the electromagnetic Casimir energy ($E_C$) in Region I in terms of the Dirichlet ($E_D$) and Neumann ($E_N$) energies,
\begin{eqnarray}
E_{C}^I(\Lambda) &=&E_{D}^I(\Lambda)+E_{N}^I(\Lambda)-\sum_{i=1}^{3} E_1(d_i,\Lambda),
\label{decomp}
\end{eqnarray}
where $E_1(d_i,\Lambda)$ is the Casimir energy in 1-dimension with $d_1 = a,\,\, d_2=b,\,\, d_3=c$.
The cutoff, $\Lambda$, reminds us that these Casimir energies are all cutoff dependent. We have already constructed the first two terms in eq.~(\ref{decomp}), and   $E_1(d_i,\Lambda)=d_i\Lambda^2/2\pi-\zeta(2)/(4\pi d_i)$ is well known.  Note that any terms that involve $\eta$ in $E_{\eta}^I$ cancel upon summation. This includes the surface divergence $\sim S\Lambda^3$, but the perimeter divergence 
$\sim L\Lambda^2$ remains. In the literature, the possibility that this term is cancelled by a similar contribution from the exterior of the parallelepiped has been suggested.
If the conducting walls have zero thickness and yet remain a perfect conductor,
this cancellation may occur. However,  we believe this to be unphysical and contrived. 
Furthermore, accounting for the energy
from the exterior of the parallelepiped is essential to determine its geometry dependence. 
In the piston geometry, all cutoff dependence\cite{divergence} drops out,
and exterior energies are handled unambiguously.

Using the methods outlined for the scalar field together with the decomposition in eq.\,(\ref{decomp}), 
we obtain an exact expression for the electromagnetic Casimir force on the  piston  with $b=c$ in the limit $h\to\infty$,
\begin{eqnarray}
F_{C}&=&-\frac{3\zeta(4)}{8\pi^2 a^4}A+\frac{\zeta(2)}{8\pi a^2}-\frac{J_C}{32\pi^2 A}
\nonumber\\
&&+\frac{\pi^2 A}{16\,a^4}\sum_{\{m,n\}}\!\!\pri\frac{\coth(f_{mn}(b/a))}{\sinh^2(f_{mn}(b/a))f_{mn}(b/a)},
\label{forceEM}
\end{eqnarray}
with $f_{mn}(x)\equiv\pi\sqrt{m^2+n^2}\,x$ and $\{m,n\}\in\mathbb{Z}^2\setminus\{0,0\}$.
The sum in the  last term is highly convergent and exponentially small as $a\to 0$.
Temperature corrections, calculated using the techniques in Ref.~\cite{Scard2}, will be presented in Ref.\,\cite{following}. The first term in eq.\,(\ref{forceEM}) is Casimir's famous result for the force between two parallel plates of area $A$ separated by a distance $a$ \cite{Casimir}.  Note that the perimeter term $\sim
P/a^3$ in eq.~(\ref{parallel}) cancelled between the two electromagnetic polarizations. The
remaining first correction, an edge effect, $ \sim 1/a^2 $, provides a repulsive contribution.
The coefficient of the $a$-independent part is $J_C\equiv J_{-1}+J_{+1}=2 Z_2(1,1;4)\approx 12.053$.

Before discussing the implications of this result, we extend it to pistons of arbitrary cross section.  
Solutions to the wave equation in a piston of arbitrary cross section factor into the product of solutions on the line $[0,a]$ and solutions in $\mathcal{S}\subset\mathbb{R}^2$, the cross section of the piston.  The density of states is therefore a convolution of the two dimensional density, $\rho_{\mathcal{S}}(\mathcal{E})$, and the density on a line, $\rho_{1}(\mathcal{E})$.
$\rho_{1}(\mathcal{E})$ is trivial and the smoothed $\rho_{\mathcal{S}}(\mathcal{E})$ can be found in Ref.~\cite{Baltes} for Dirichlet and Neumann BC. A decomposition of the EM case into Dirichlet and Neumann sub-problems, similar to eq.~(\ref{decomp}), exists (for details, see Ref.\,\cite{following}).  The results for scalar and EM fields are
\begin{eqnarray}
\label{eq:generalEM}
F_{\eta}&=&  - \frac{3\zeta(4)}{16\pi^2 a^4}A - \eta\frac{\zeta(3)}{32\pi a^3}P - \frac{\zeta(2)\chi}{4\pi a^2} + \mathcal{O}(1),\nonumber\\
F_{C}&=&-\frac{3\zeta(4)}{8\pi^2
a^4}A+\frac{\zeta(2)(1-2\chi)}{4\pi a^2} +\mathcal{O}(1).
\end{eqnarray}
Here $\chi$ depends on the 2-dimensional cross section, as
\begin{equation}
\chi\equiv\sum_{i}\frac{1}{24}\left(\frac{\pi}{\alpha_i}-\frac{\alpha_i}{\pi}\right)
+\sum_j\frac{1}{12\pi}\int_{\gamma_j}\kappa(\gamma_j)d\gamma_j,
\end{equation}
where $\alpha_i$ is the interior angle of each sharp corner and
$\kappa(\gamma_j)$ is the curvature of each smooth section described by the curve $\gamma_{j}$. For example, $\chi=1/4$ for a rectangle and $\chi=1/6$ for all smooth shapes.   Substituting $\chi=1/4$ we find the first 
three terms of eq.~(\ref{parallel}) and the first two terms of eq.~(\ref{forceEM}). 
The $a$-independent terms in eqs.~(\ref{parallel}) and (\ref{forceEM}), denoted by $-J/A$, 
are difficult to recover for arbitrary cross sections with the methods used here.

We now discuss the implications of our results for issues of attraction versus repulsion.  The term proportional to $-J_C/A$ in eq.~(\ref{forceEM}) plays an important part.   To understand its origin consider the case $h\gg b\gg a$.  In this regime the cutoff independent contribution to $F_{C}$ from Region I contains \emph{no $a$-independent term}.   
The terms that scale like $1/a^{4}$ and $1/a^{2}$ remain.  They come from paths whose lengths vanish as $a\to 0$.  
The exponentially small terms come from paths that hit both sides of the cavity.  
Their contributions vanish like $e^{-2\pi b/a}$, since the minimum wave number in the vertical direction, 
$\sim 1/a$,  acts like a mass for horizontal propagation. The $-J_C/A$ term in eq.~(\ref{forceEM}) is instead the sole 
cutoff independent relic of Region II when $h\gg b\gg a$.  It comes entirely from optical paths lying in the horizontal 
plane, such as Fig.~\ref{2dnew}(i).  These contributions are extensive in the vertical direction, and thus depend on 
$a$ as $\sim - (h-a)/b^{2}$.  This repulsion between the top and bottom plates of Region II leads to the $a$-independent attraction in eq.~(\ref{forceEM}).  All other optical paths in Region II give contributions that vanish as $h\to \infty$.  Now consider Region I when $a\gg b$.  
Following the same logic, an  extensive contribution to the energy, $\sim - a/b^{2}$, now arises from horizontal paths in 
Region I.  This cancels the contribution from horizontal paths in Region II leaving a Casimir force, $F_{C}$, 
that approaches zero (from below) exponentially fast as $a/b\to\infty$.  This can easily be understood:  
In this limit the horizontal wave number, $\sim 1/b$, now plays the role of a mass for propagation in the vertical direction, 
which resembles a 1-dimensional system.   This cancellation occurs for any cross sectional shape, 
differing only by the value of $J_C$. 

\begin{figure}[t]
\begin{center}
\scalebox{0.68}{\includegraphics{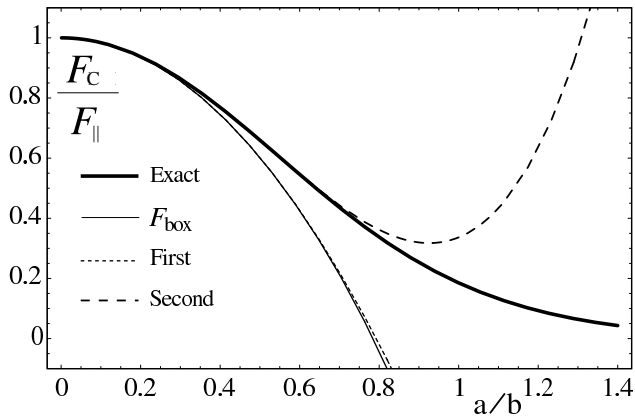}}
\caption{\label{fig:ForceEM}The force on a 3-dimensional piston
$F_{C}$ with a square cross section due to EM quantum fluctuations, as a function of $a/b$
(see Fig.~\ref{2dnew}), normalized to the parallel plates result
$F_{\parallel}=-3\zeta(4)A/(8\pi^2 a^4)$. The exact expression is the thick
solid line, and the first and second corrections are the dashed
lines. The thin solid line is the force $F_{\rm box}$ that comes
from differentiating the finite part of the energy in a single
box.}
\end{center}
\end{figure}

Had we not formed the piston, but only considered the finite part of the energy in  the  ``box'' defined by Region I alone, 
the term $-J_C/(32\pi^{2}\!A)$ would have been absent from the expansion equivalent to eq.~(\ref{forceEM}).  
As a result $F_{\rm box}\to+J_C/(32\pi^2\!A)$ for $a\gg b$. In particular, one can show $F_{\rm box}>0$, for $a/b>0.785$.  
This explains the claim of repulsion that has appeared in the literature:  Without Region II 
(or some open region that allows rigid motion of the partition) the Casimir energy of the parallelepiped is, 
in fact, cutoff dependent.  If the cutoff dependence is somehow ignored, a repulsive force at large $a/b$ 
remains as an artifact.  In Fig.~\ref{fig:ForceEM},
the force is plotted  normalized to the parallel plates result
$F_{\parallel}=-3\zeta(4)A/(8\pi^2 a^4)$. The solid line gives our exact
result, and the dashed lines are the results from successively
including each of the finite corrections of Eq.\,(\ref{forceEM}).
There is good agreement for $a\lesssim 0.3 b$ with the first correction,
and for $a\lesssim 0.7 b$ upon inclusion of the second (constant) term.

In summary, we have computed the Casimir force acting on a rectangular piston (Fig.~\ref{2dnew}) exactly for both scalar and EM fields --- one of a rather small set of exact results for physically interesting geometries.  
We make quantitative predictions for the forces in this physically realizable system (Fig.~\ref{fig:ForceEM}), which are found to be attractive for any rectangular cross section.
Casimir force calculations for isolated parallelepipeds that appear to give repulsion have ignored cutoff   dependence\cite{divergence}. The simplest way to cancel the cutoff dependence is to introduce Region II, which is found to also cancel a cutoff {\it independent} part of the energy.  This changes the repulsion into attraction, and is expected to occur for any cross section. 

The corrections to the leading parallel plate force  
can be understood by analyzing an optical path representation of the Green's function.  For a scalar field the first correction is one power of $a$ suppressed, and comes from paths that reflect only once from the piston's sides.  It is repulsive for Dirichlet and attractive for Neumann BC.  For the electromagnetic field this correction cancels between Dirichlet and Neumann modes.  The first non-zero correction is two powers of $a$ suppressed and comes from paths that reflect once off an edge. It is repulsive.  

We also obtain the leading correction to the parallel plate result for arbitrary shape of the 
 cross section.  This correction is quite sensitive to the geometry of the base; 
 it is the same for all smooth curves, but distinct for shapes with sharp edges. 
Note that the expansion parameter $a$ provides a length scale for smoothing 
irregularities of the shape at small scales.

\vspace*{2ex}
We thank L.~Levitov for discussions.  M.~P.~H., R.~L.~J., and A.~S. are supported in part by funds provided by the U.S.~Department of
Energy (D.O.E.) under cooperative research agreement DE-FC02-94ER40818.
M.~K. is supported by NSF grants  DMR-04-26677 and PHY99-07949.  A.~S.~acknowledges support from Bruno Rossi and John A.~Whitney Scholarships.

\end{document}